%% file: average_prosparse_Final.tex
\documentclass[journal]{IEEEtran}

\include{commands}

\usepackage[nosort]{cite}

\usepackage{xcolor}
\usepackage{soul}

\usepackage{mathtools}
\mathtoolsset{showonlyrefs=true} 

\providecommand{\floor}[1]{\lfloor #1 \rfloor}

\newcommand{\PP}{\mathbf{P}}
\DeclareMathOperator{\coef}{coef}
\renewcommand{\l}{\ell}

\begin{document}

\title{Sparse Representation in Fourier and Local Bases Using ProSparse: A Probabilistic Analysis}
\author{
Yue~M.~Lu,~\IEEEmembership{Senior~Member,~IEEE}, 
Jon~O\~{n}ativia,~\IEEEmembership{Member,~IEEE},
and Pier~Luigi~Dragotti,~\IEEEmembership{Fellow,~IEEE}
\thanks{Y. M. Lu is with the John A. Paulson School of Engineering and Applied Sciences, 
Harvard University, Cambridge, MA 02138, USA (e-mail: yuelu@seas.harvard.edu). His work was supported in part by the US National Science Foundation under grant CCF-1319140, and by the US Army Research Office under contract W911NF-16-1-0265.}
\thanks{J. O\~{n}ativia was with the Department of Electrical and Electronic Engineering, 
Imperial College London, London SW7 2AZ, UK. He is now with Egile, 20850, Spain (e-mail: jon.onativia@ieee.org). P. L. Dragotti is with the 
Department of Electrical and Electronic Engineering, 
Imperial College London, London SW7 2AZ, UK 
(e-mail: p.dragotti@imperial.ac.uk). J. Onativia and P. L. Dragotti were supported by the European Research Council Starting Investigator Grant through the RecoSamp Project under Award 277800.}
\thanks{This work was presented in part at the Signal Processing with Adaptive Sparse Structured Representations (SPARS 2015) Workshop.}%
}

\markboth{}{Lu \MakeLowercase{\textit{et al.}}: Sparse Representation in Fourier and Local Bases Using ProSparse}

\maketitle

\begin{abstract}
Finding the sparse representation of a signal in an overcomplete dictionary has attracted a lot of attention over the past years. This paper studies ProSparse, a new polynomial complexity algorithm that solves the sparse representation problem when the underlying dictionary is the union of a Vandermonde matrix and a banded matrix. Unlike our previous work which establishes deterministic (worst-case) sparsity bounds for ProSparse to succeed, this paper presents a probabilistic average-case analysis of the algorithm. Based on a generating-function approach, closed-form expressions for the exact success probabilities of ProSparse are given. The success probabilities are also analyzed in the high-dimensional regime. This asymptotic analysis characterizes a sharp phase transition phenomenon regarding the performance of the algorithm.
\end{abstract}

\section{Introduction}

Let $\vy \in \C^N$ be a complex signal that admits a sparse representation in an overcomplete 
dictionary. That is, 
\begin{equation}\label{eq:model}
\vy = \mD \, \vx, 
\end{equation}
where $\mD = [\mPsi, \mPhi]$ is the union of two bases or frames and $\vx$ is a sparse coefficient vector. 
Given $\vy$, the sparse vector $\vx$ can be recovered by greedy algorithms such as orthogonal matching pursuit and its variants (e.g.,~\cite{MallatZ:93, Pati:1993, DaiM:2009, NeedellT:2009}) or by convex relaxation techniques such as basis pursuit (BP) \cite{ChenDS:98,EladB:02}. The performance of these algorithms and the corresponding requirements on the dictionary $\mD$ are now well understood \cite{EladB:02, GribonvalN:2003, FeuerN:2003, Tropp:04, CandesR:2005, Tropp:2008, Kuppinger:2012, Studer:2012, Pope:2013}. For a comprehensive overview of this topic, we refer the reader to the book~\cite{Elad:2010}.



Recently, a new algorithm, named \emph{ProSparse} in \cite{DragottiL:2014}, has been presented for the case where one of the sub-dictionaries, $\mPsi \in \C^{N \times M}$, is a Vandermonde matrix (with $M \ge N$), and the other sub-dictionary $\mPhi$ is an $N \times N$ banded matrix. We note that a canonical example of $\mPsi$ is the Fourier basis (or frame), whereas typical examples of $\mPhi$ include the standard basis ($\mPhi = \mI_N$), the short-time Fourier transform, and pulse signals that are often encountered in radar and ultrasound applications. It was shown in \cite{DragottiL:2014} that ProSparse, an algorithm with polynomial complexity, can recover $\vx$ from \eref{model}, provided that 
\begin{equation}
\label{eq:det_bound}
2(P + b_{\mPhi})(K+\tau) < N+\tau (2b_{\mPhi}+1).
\end{equation}
Here, $P$ and $K$ denote the sparsity level of $\vx$ in $\mPsi$ and $\mPhi$, respectively: the overall sparsity of $\vx$ is then equal to $P + K$; $b_{\mPhi}$ denotes the bandwidth of the banded matrix $\mPhi$; and $\tau$ is a constant that takes two possible values: $\tau = 0$ if $\mPsi$ is a circulant matrix and $\tau = 1$ otherwise. { It is interesting to point out that the performance bound in \eref{det_bound} does not depend on $M$, the size of the sub-dictionary $\mPsi$. This is in fact a consequence of a useful property of ProSparse: unlike BP, the performance of ProSparse does not degrade when $\mPsi$ is highly overcomplete, \emph{i.e.}, when $M$ is much larger than $N$. (More discussions on this point can be found in \sref{comparison}.)}

For the prototypical case where $\mPsi$ is the Fourier basis and $\mPhi = \mI_N$, we have $\tau = 0$ and $b_{\mPhi} = 0$. The bound in \eref{det_bound} can then be simplified as
\begin{equation}\label{eq:bound_Dirac}
P K < N/2. 
\end{equation}
We note that this sparsity requirement is much weaker than the corresponding tight BP bound \cite{EladB:02} and the unicity bound \cite{ChenDS:98} in the literature (see Figure~1 in~\cite{DragottiL:2014} for a comparison of these different bounds.) It also improves the theoretical recovery threshold given in \cite[Theorem 7]{Studer:2012} (when specialized to the case of Fourier and identity matrices) by a factor of two. Interestingly, the same sparsity requirement \eref{bound_Dirac} was previously established in an early work \cite{DonohoS:89}, under the assumption that the support pattern of $\vx$ in the sub-dictionary $\mPhi$ is \emph{known}. ProSparse achieves the same performance bound \emph{without} the knowledge of the support of $\vx$. Finally, we also point out that \eref{bound_Dirac} still suffers from the so-called \emph{squared-root bottleneck} \cite{Tropp:2008}.

As a deterministic and worst-case bound, the condition in \eref{det_bound} guarantees the success of ProSparse in recovering every single $\vx$ for which it holds. The bound is also tight in the sense that one can construct counterexamples for which \eref{det_bound} is not satisfied and the algorithm fails. However, these counterexamples have special support structures, the occurrence of which can be rare. Motivated by our desire to understand the typical performance of ProSparse, we present in this paper an average-case analysis of ProSparse within a probabilistic setting. Our contributions are as follows:

1. \emph{Exact success probabilities.} We provide closed-form expressions for the exact success probabilities of ProSparse in recovering a sparse signal $\vx$, when the sparsity pattern of $\vx$ is drawn uniformly at random. This model turns out to be equivalent to the problem of discrete circle covering \cite{Holst:1985,Barlevy:2015} in geometric probability. The success probabilities of ProSparse can be evaluated by known formulas in the literature, but we provide an alternative expression in Proposition~\ref{prop:exact_proba}. Based on a simple generating-function approach, our new expression can be efficiently evaluated via the Fourier transform, and its simple form also plays a key role in our subsequent analysis in the high-dimensional setting. 

2. \emph{Asymptotic analysis.} As the main technical contribution of this work, we analyze the above-mentioned success probabilities in the asymptotic, high-dimensional regime. Our asymptotic analysis reveals a phase transition phenomenon regarding the performance of ProSparse. {Specifically, suppose that the support of $\vx$ involves $K = \floor{\alpha N^\delta}$ atoms from $\mPhi$ and $P = \floor{\beta N^{1-\delta} \log N}$ atoms from $\mPsi$. Here, $0 < \delta \le 1$, $\alpha > 0$ and $\beta > 0$ are three positive constants. (When $\delta = 1$, the constant $\alpha$ is further required to be less than 1.) We show in Proposition~\ref{prop:main_result} that, if the $K$ atoms from $\mPhi$ are chosen uniformly at random, then}
\begin{equation}\label{eq:phase}
\PP(\text{ProSparse recovers } \vx) \xrightarrow[]{N\rightarrow \infty} \begin{cases}
1, &\text{if } \beta < \beta_c(\alpha, \delta)\\
0, &\text{if } \beta > \beta_c(\alpha, \delta),
\end{cases}
\end{equation}
where the critical threshold is given by
\begin{equation}\label{eq:phase_bound}
\beta_c(\alpha, \delta) = \begin{cases}
\frac{\delta}{2 \alpha}, &\text{if } 0 < \delta < 1\\
\frac{-1}{2 \log(1-\alpha)}, &\text{if } \delta = 1.
\end{cases}
\end{equation}

\begin{figure}[t]
\centering
\subfigure[$K = \floor{\alpha N}$]{\label{fig:phase_transition_1}
\includegraphics[width=0.22\textwidth]{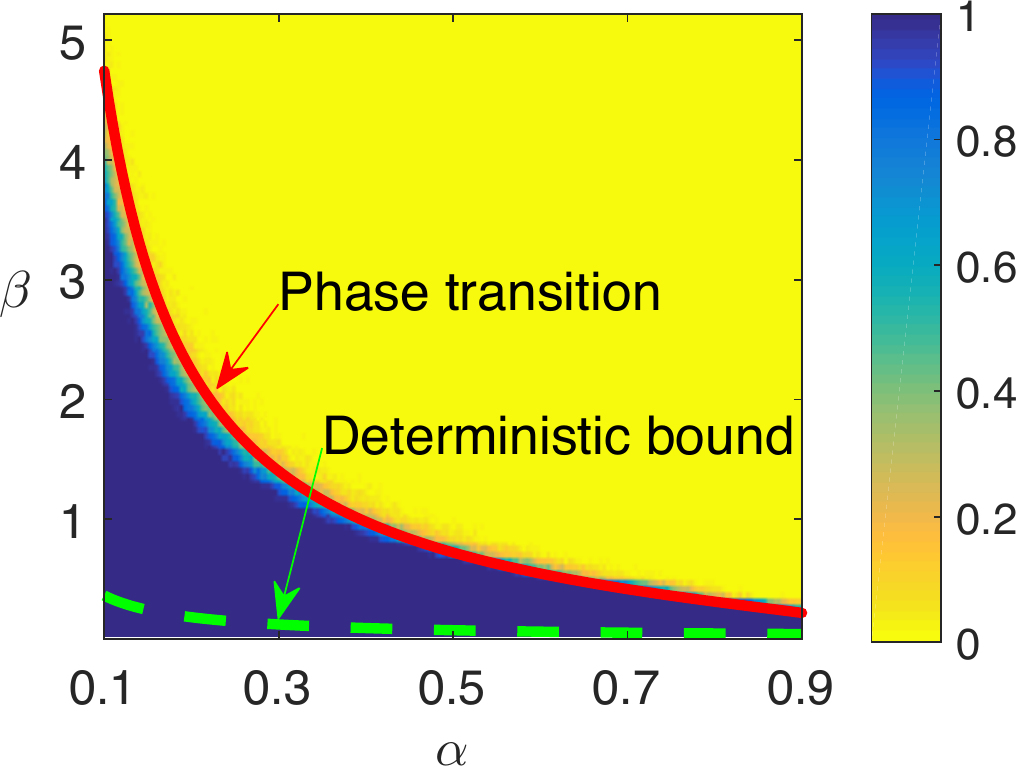}\vspace{1ex}
}
\subfigure[$K = \floor{\alpha N^{0.6}}$]{\label{fig:phase_transition_2}
\includegraphics[width=0.22\textwidth]{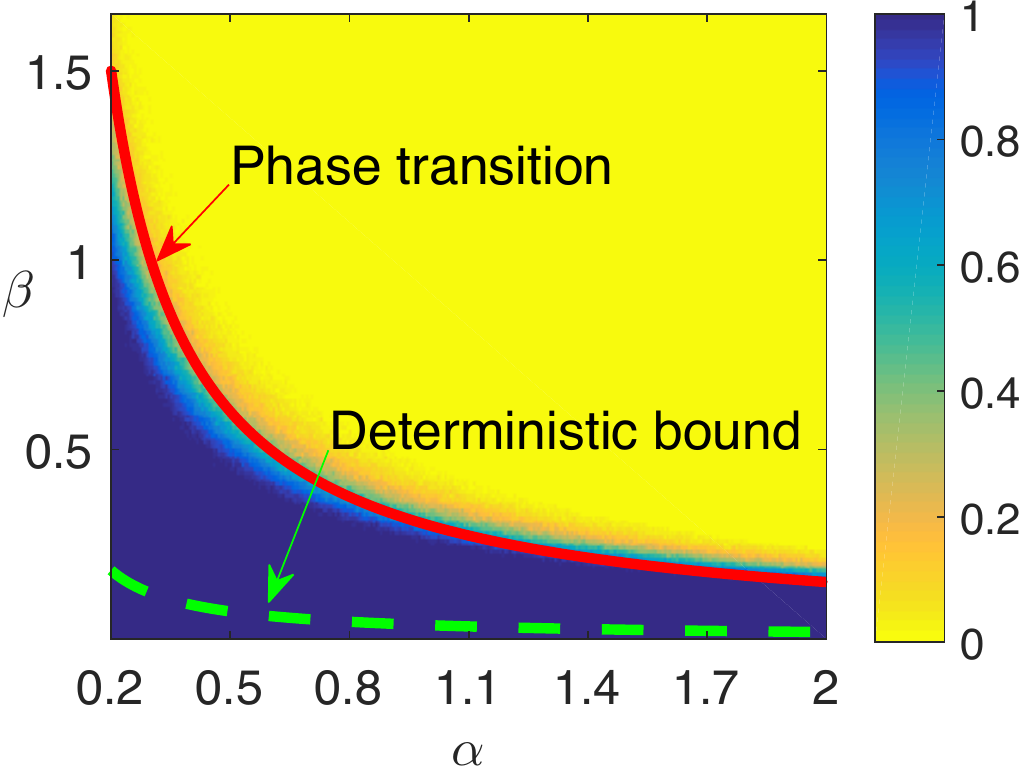}\vspace{1ex}
}
\vspace{-1ex}
\caption{Success probabilities of ProSparse: simulations vs. theory. The figures show empirical success probabilities of ProSparse in recovering a sparse signal when the dictionary is the union of a Fourier frame and the canonical basis.  The signal dimension is set to $N = 10^6$. In (a), the number of Fourier atoms is $P = \floor{ \beta \log N}$, and the number of ``spikes'' is $K = \floor{\alpha N}$. A different scaling is shown in (b), where $P = \floor{\beta N^{0.4} \log N}$ and $K = \floor{\alpha N^{0.6}}$. For each pair of $(\alpha,\beta)$, 100 different spike patterns have been generated uniformly at random. In both (a) and (b), the solid lines show the predicted phase transition boundaries given in \eref{phase_bound}, and the dashed lines show the deterministic bound given in \eref{det_bound}.}
\label{fig:phase_transition}
\end{figure}


We illustrate this phase transition phenomenon in \fref{phase_transition}, for two different values of $\delta$. The two axes of the figures correspond to the parameters $\alpha$ and $\beta$, respectively. The red, solid lines show the phase transition boundaries [$\beta = \frac{-1}{2 \log(1-\alpha)}$ for $\delta = 1$ in \fref{phase_transition_1} and $\beta = \frac{\delta}{2 \alpha}$ for $\delta = 0.6$ in \fref{phase_transition_2}]. The theoretical predictions match very well with results from Monte Carlo simulations (for $N = 10^6$), delineating the two regions where ProSparse succeeds or fails with high probability. For comparison, we also plot the deterministic bound given in \eref{det_bound}, which is much more pessimistic than the actual average-case performance of the algorithm.

The rest of the paper is organized as follows. \sref{overview_prosparse} provides a brief overview of ProSparse, highlighting its key ideas. In \sref{exact_proba}, we present an average-case performance analysis of ProSparse by deriving closed-form expressions for the exact success probabilities of ProSparse in sparse recovery. \sref{phase_trans} studies the asymptotic regime, where we prove the phase transition result shown in \eref{phase} and compare it with related results in the literature. We conclude in Section~\ref{sec:conclusion}.

\section{Overview of ProSparse}
\label{sec:overview_prosparse}

In this section, we briefly recall the key ideas behind ProSparse \cite{DragottiL:2014}. This discussion serves as the basis for our probabilistic performance analysis carried out in later sections. 

Let $\mD=[\mPsi, \mPhi]$ be the union of two sub-dictionaries. Of the two, $\mPsi \in \C^{N \times M}$ is a Vandermonde matrix with entries
\[
\Psi_{n, m} = \xi_m^{n},\ \text{ for } 0 \le n < N, 0 \le m < M,
\]
where $\set{\xi_m \in \C: 0 \le m < M}$ is a set of distinct numbers. For simplicity, we assume that the other sub-dictionary $\mPhi$ is an $N \times N$ identity matrix. (The more general case involving banded matrices can be handled by following the same idea. See Remark~\ref{rem:banded} below for discussions.) Let $\vx_1 \in \C^M$ and $\vx_2 \in \C^N$ be two vectors containing $P$ and $K$ nonzero entries, respectively. Our goal is to recover the $(P, K)$-sparse vector $\vx = [\vx_1^T, \vx_2^T]^T$ from the measurement $\vy = [\mPsi, \mPhi] \, \vx$. 

%
%
%

Given the specific structure of the two sub-dictionaries, the $n$th entry of $\vy$ can be written as follows:
\begin{equation}
y_n = 
\sum_{p=1}^{P} a_p \, \xi_{m_p}^n + \sum_{k=1}^{K} b_k \, \delta_{n - n_k},
\qquad n = 0,1,\ldots,N-1.
\end{equation}
Here, $\delta_n$ is the Kronecker delta function; the indices $\set{m_p}_{p=1}^{P}$ and $\set{n_k}_{k=1}^{K}$ specify the sparsity patterns of $\vx_1$ and $\vx_2$, respectively; $\set{a_p}$ and $\set{b_k}$ denote the corresponding nonzero coefficients. 

ProSparse is based on two simple observations:

First, since the basis elements from $\mPhi$ are local, many of the entries in $\vy$ are only due to the Vandermonde components $\mPsi \vx_1$. Specifically, we can find ``clean'' windows of consecutive entries of $\vy$ that can be expressed as merely a sum of exponentials
\begin{equation}
\label{eq:y_exps}
y_n  = \sum_{p=1}^{P} a_p  \xi_{m_p}^n,
\end{equation}
where we have restricted the index $n$ to a ``clean'' window. See \fref{gaps} for an illustration.

\begin{figure*}[t]
\centering
\includegraphics[width=0.7\linewidth]{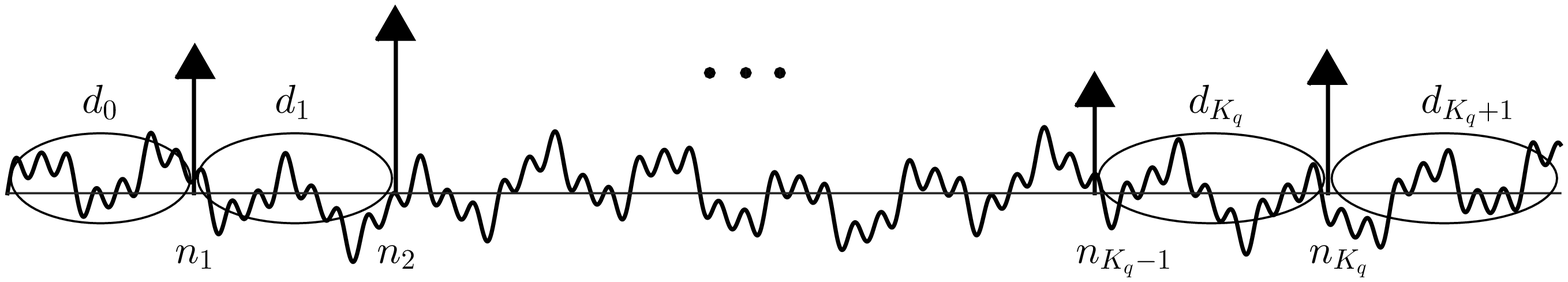}
\caption{Samples of $y_n$ located between consecutive spikes are only due to the Vandermonde components, and they can be written in the form of \eref{y_exps}. The sum of all the gaps $\set{d_k}$ is equal to $N - K$. ProSpare sequentially searches through all sliding windows of size $2P$. It is able to recover the unknown parameters $\set{a_p, \xi_{m_p}}_p$ in \eref{y_exps} [and subsequently the sparse signal $\vx$] if there is a gap of length at least $2P$.}
\label{fig:gaps}
\end{figure*}

Second, as long as we have at least $2P$ consecutive entries of the form \eref{y_exps}, we can exactly recover the unknown parameters $\set{a_p, \xi_{m_p}}_{p=1}^{P}$ from these entries. A classical approach is to use Prony's method, first proposed by Baron de Prony in 1795 \cite{Prony:1795}. Since then, this algebraic method for parameter estimation has been used in many different contexts \cite{StoicaM:97}. 

Exploiting the two observations stated above, the ProSparse algorithm operates by performing an exhaustive search over all possible sliding windows $y_{\l}, y_{\l+1}, \ldots, y_{\l+2P-1}$ of length $2P$, for every $P < N/2$. Note that, since these sliding windows are sequential, the exhaustive search has polynomial complexity. For each candidate sliding window, the algorithm uses Prony's method to try to estimate the parameters $\set{a_p, \xi_{m_p}}_{p=1}^{P}$, from which we can build a residual vector $\vr$ with entries $r_n = y_n - \sum_{p=1}^{P} a_p \xi_{m_p}^n$. If the residual is sparse, it then directly corresponds to the spikes $\sum_{k=1}^{K} b_k \, \delta_{n - n_k}$. When the sparsity levels satisfy \eref{det_bound} (with $\tau = 0$ and $b_{\mPhi} = 0$), we can show that there always exist clean intervals of length at least $2P$, and thus ProSparse is guaranteed to recover the original signal $\vx$. We refer the reader to \cite[Proposition~2]{DragottiL:2014} for a formal proof. Strategies for further reducing computational complexity and discussions on issues involving multiple solutions can also be found in \cite{DragottiL:2014}.

\begin{remark}[The Fourier basis]\label{rem:Fourier}
Consider the special case when $\mPsi$ is the Fourier basis, that is, when $M = N$ and $\Psi_{n, m} = \omega^{mn}$ with $\omega = e^{-2\pi j / N}$. The problem becomes circulant due to the  periodicity of the exponential function $\omega^n$. This means that the search for the clean windows can be performed on a ring, with the last entry $n=N-1$ immediately followed by the first entry $n=0$. The parameter $\tau$ in the general bound \eref{det_bound} is introduced to exactly take into account this special case: $\tau = 0$ if the Vandermonde matrix $\mPsi$ is the Fourier matrix, and $\tau = 1$ for other generic Vandermonde matrices for which we cannot exploit the periodicity.
\end{remark}

\begin{remark}[Banded matrices] \label{rem:banded} The idea described above is applicable to more general cases when $\mPhi$ is a banded matrix, \emph{i.e.},
\[
\Phi_{n, m} = 0 \quad \text{if} \quad m < n - b_{\mPhi} \quad \text{or} \quad m > n + b_{\mPhi},
\]
where $b_{\mPhi}$ is the bandwidth of $\mPhi$. Let $\set{n_k: 1 \le k < K}$ denote the indices corresponding to the sparsity pattern of $\vx_2$. These indices break the interval $[0, N-1]$ into $K + 1$ segments (or $K$ segments if $\mPsi$ is periodic.) It is easy to see that a sufficient condition for ProSparse to work is to have at least one segment whose length is greater than or equal to $2P + 2b_{\mPhi}$. This is the origin of the first term on the left-hand side of \eref{det_bound}.
\end{remark}

\begin{remark}[Further generalizations] Throughout this paper, we assume that $\mPsi$ is a Vandermonde matrix. However, it is possible to loosen this requirement by considering a larger class of $\mPsi$-matrices satisfying the so-called \emph{local sparse reconstruction property} \cite{DragottiL:2014}. In short, those are matrices that allow for the efficient reconstruction of sparse signals from \emph{any} small blocks of consecutive elements in the transform domain. Beyond Vandermonde matrices considered in this work, other examples of such matrices include the DCT transform (and its variants) and random Gaussian dictionaries. More details on this possible generalization, including the precise definition of the local sparse reconstruction property, can be found in \cite[Sec. IV]{DragottiL:2014}.
\end{remark}

Finally, we note that the bounds in \eref{det_bound} and \eref{bound_Dirac} are essentially tight: we can construct sparse signals for which the bounds do not hold and ProSparse fails to recover the signals. One such counterexample is the ``picket-fence'' signal, with the spikes (or local atoms) from $\mPhi$ placed at equally-spaced locations. However, such signals are rare, if the sparsity patterns of $\vx$ are drawn uniformly at random. The next section provides an exact probability analysis of the performance of ProSparse. We will show that, indeed, the deterministic bounds \eref{det_bound} and \eref{bound_Dirac} are overly conservative.

\section{Exact Probabilistic Performance Analysis}
\label{sec:exact_proba}

From the discussions in the previous section, we see that the success of ProSparse crucially depends on whether there exists a gap of sufficient length after we break the interval $[0, 1, \ldots, N-1]$ into a given number of pieces. In what follows, we study this question in a probabilistic setting, with randomly chosen ``breaking points''.

\subsection{Probability Model and Analysis}

Consider a set of $N$ integers $0, 1, \ldots, N-1$ arranged along a line. We draw $K$ distinct numbers, denoted by $n_1, n_2, \ldots, n_K$, by using random sampling without replacement from the set. Now consider the order statistics obtained by arranging the $n_k$'s in increasing order:
\begin{equation}\label{eq:n_order}
0 \le n_{[1]} < n_{[2]} < \ldots < n_{[K]} \le N-1.
\end{equation}
We define the gaps between consecutive points as
\begin{equation}\label{eq:gaps}
d_k \bydef n_{[k+1]} - n_{[k]} - 1, \quad \text{for } 0 \le k \le K,
\end{equation}
where two additional end-points $n_{[0]} = -1$ and $n_{[K+1]} = N$ are introduced. By construction, $d_k \ge 0$, and it must hold that
\begin{equation}\label{eq:sum_constraint}
\sum_{k=0}^{K} d_k = N - K.
\end{equation}
Discrete random variables $\set{d_k}$ satisfying the above sum constraint arise in statistical physics, and they are often referred to as the Bose-Einstein model; see \cite[pp. 20--21]{Feller:1968}.

\fref{gaps} illustrates a typical realization of this probabilistic model. We note that, in our analysis of ProSparse, the numbers $\set{n_k}$ represent the support of a $K$-sparse signal whose sparsity pattern is chosen at random, whereas the gaps $\set{d_k}$ are the lengths of those clean intervals within which \eref{y_exps} holds.

Denote the maximum gap by
\begin{equation}
\mathit{\mathit{\Delta}}_{N, K} \bydef \max_{0 \leq k \leq K} \lbrace d_k \rbrace.
\label{eq:max_gap}
\end{equation}
A central problem in our analysis is to understand the probability distribution of $\mathit{\mathit{\Delta}}_{N, K}$, \emph{i.e.}, to compute
\[
h_{N, K}(s) \bydef \PP(\mathit{\mathit{\Delta}}_{N, K} < s)
\]
for any positive integer $s$. For the special case when $\mPsi$ is the unitary Fourier matrix (see Remark~\ref{rem:Fourier} in \sref{overview_prosparse}), we need to modify $\mathit{\mathit{\Delta}}_{N, K}$ and study the probability distribution of a related quantity
\begin{equation}
\mathit{\mathit{\Gamma}}_{N, K} \bydef \max \Big\lbrace d_0+d_{K}, \max_{1 \leq k < K} \lbrace d_k \rbrace  \Big\rbrace,
\label{eq:max_gap_circ}
\end{equation}
which takes into account the circulant nature of the problem.

Note that the model we consider here is the discrete version of the classical spacing problem (see, \emph{e.g.}, \cite{Barton:1956, Flatto:1962, Pyke:1965, Holst:1980}), in which a unit circle is broken at $K$ randomly chosen points and the question of interest is to study the various statistical properties of the lengths of the intervals between consecutive points. Compared to the long history and vast literature of the continuous spacing problem, discussions on the discrete version of the problem appeared in the literature much later \cite{Holst:1985, Drakakis:2009, Huillet:2011,Barlevy:2015}. It was first introduced by Holst \cite{Holst:1985}, who computed, among other quantities, the probability distribution of the following random variable\footnote{We have adapted the definition of $V_{N, K}$ and the subsequent formula \eref{g_v} to reflect the fact that we are using slightly different notational conventions from those in \cite{Holst:1985,Barlevy:2015}.}
\[
V_{N, K} \bydef \sum_{k=0}^K \left(d_k - (s-1)\right)_+,
\]
where $(x)_+ = \max\{x, 0\}$. Clearly, the maximum gap $\mathit{\mathit{\Delta}}_{N, K} < s$ if and only if $V_{N, K} = 0$. It then follows that we can use the probability distribution of $V_{N, K}$ presented in \cite{Holst:1985} (see \cite[p. 123]{Barlevy:2015} for a correction) to get
\begin{equation}\label{eq:g_v}
h_{N, K}(s) = \PP(V_{N, K} = 0) = \frac{\sum_{\ell=0}^{K+1} (-1)^\ell \binom{K+1}{\ell} \binom{N - \ell s}{K}}{\binom{N}{K}}.
\end{equation}
As an alternative way to reach \eref{g_v}, we can interpret $h_{N, K}(s)$ as the distribution of the largest order statistic of the random variables $\set{d_0, d_1, \ldots, d_K}$. Although $d_i$ are not independent, they form an exchangeable family. Thus, \eref{g_v} can also be seen as a special case of a general recursive formula for the distribution of orders statistics of exchangeable random variables (see, \emph{e.g.}, \cite[p. 46]{DavidN:03}).


The right-hand side of \eref{g_v} involves an alternating sum of products of binomial coefficients. Therefore, evaluating \eref{g_v} numerically becomes challenging even for moderate values of $N$ and $K$. In what follows, we prove an alternative expression for $h_{N, K}(s)$, which, to our knowledge, has not been presented in the literature before. Based on integer powers of a certain generating function, our new expression can be efficiently and accurately evaluated by the discrete Fourier transform. More importantly, this new expression plays a key role in the asymptotic analysis of $h_{N, K}(s)$ to be carried out in \sref{phase_trans}.

Before presenting our result, we first introduce the notation
\begin{equation}
\coef\left\{(c_0+c_1\,x+\ldots+c_n\,x^n), x^\ell\right\} = c_\ell,
\end{equation}
to refer to the $\ell$th coefficient of a polynomial. (Note that $c_\ell \equiv 0$ for $\ell > n$.) 

\begin{proposition} \label{prop:exact_proba}
For any positive integer $s$,
\begin{equation}
h_{N,K}(s) = \frac{\coef\left\{f_s(x)^{K+1}, x^{N-K}\right\}}{\binom{N}{K}},
\label{eq:exact_prob}
\end{equation}
where $f_s(x) \bydef \sum_{i = 0}^{s-1} x^i$ is a polynomial of degree $s-1$.
\end{proposition}

\begin{IEEEproof}
Consider the experiment of drawing $K$ unique numbers from $0, 1, \ldots, N-1$ by sampling without replacement. We are interested in computing the probability that the maximum gap between consecutive numbers is smaller than $s$. This probability is given by the ratio between the number of outcomes with $\mathit{\mathit{\Delta}}_{N, K} < s$ and the total number of possible outcomes. The denominator is clearly given by the binomial coefficient $\binom{N}{K}$. For the numerator, we note that the number of outcomes associated with $\mathit{\Delta}_{N, K} < s$ is equal to the cardinality of the following set
\begin{equation}\label{eq:set_d}
\Big\{\left(d_0,d_1,\ldots,d_{K}\right) \in \{0,1,\ldots,s-1\}^{K+1}: \sum_{k=0}^{K} d_k = N-K\Big\}.
\end{equation}
This set enumerates all possible configurations of gaps such that the maximum gap is always smaller than $s$, subject to the additional sum constraint \eref{sum_constraint}. This counting problem can be solved by a generating function approach: consider the polynomial
\begin{equation}\label{eq:generating}
(1 + x + x^2 + \ldots + x^{s-1})^{K+1}.
\end{equation}
It is easy to verify that the cardinality of the set is exactly equal to the $(N-K)$th coefficient of the above polynomial.
\end{IEEEproof}

\begin{remark}
Using the equivalence between polynomial multiplication and convolution, the numerator in \eref{exact_prob} can be efficiently computed by two applications of the discrete Fourier transform (DFT): one forward DFT to compute the Fourier transform of $f_s(x)$; and an inverse DFT to obtain the $(N-K)$th coefficient of $f_s(x)^{K+1}$. 
\end{remark}
\begin{remark}
The generating-function formula given in \eref{exact_prob} also has an interesting probabilistic interpretation, which is what initially led us to considering \eref{generating}. To count the number of elements in the set \eref{set_d}, we first construct $K+1$ independent random variables $d_0, d_1, \ldots, d_K$, each of which is uniformly distributed on $\set{0, 1, \ldots, s-1}$. (Note that this distribution is introduced for counting purposes; it is different from the actual distribution of the gaps $\set{d_k}$, which have to satisfy \eref{sum_constraint} and are thus \emph{not} independent of each other.) Since all configurations are equally likely, the cardinality of \eref{set_d} is equal to $s^{K+1} \PP(\sum_k d_k = N-K)$. The probability generating function of $d_k$ (for any $k$) is
\[
g(x) = \sum_{i=0}^{s-1} \PP(d_k = i) x^i = f_s(x) / s.
\]
Since the random variables are i.i.d., the probability generating function of their sum $\sum_k d_k$ is then the $(K+1)$th power of $g(x)$.
\end{remark}

Next, we consider the probability distribution of $\mathit{\Gamma}_{N, K}$, the maximum gap in the circulant setting. As shown in the following corollary, the distribution of $\mathit{\Gamma}_{N, K}$ can be obtained from that of $\mathit{\Delta}_{N, K}$.

\begin{corollary}[The circular case]
\label{corr:exact_proba_circ}
For any positive integer $s$,
\begin{equation}
\PP(\mathit{\Gamma}_{N, K} < s) = \PP(\mathit{\Delta}_{N-1, K-1} < s) = h_{N-1, K-1}(s).
\label{eq:exact_prob_circ}
\end{equation}
\end{corollary}
\begin{IEEEproof}
The proof is based on a simple conditioning argument. Due to the circulant structure of the problem, we have
\[
\PP(\mathit{\Gamma}_{N, K} < s \mid n_1 = n) = \PP(\mathit{\Delta}_{N-1, K-1} < s) = h_{N-1, K-1}(s).
\]
Since the above conditional probability does not depend on $n$,
\[
\begin{aligned}
\PP(\mathit{\Gamma}_{N, K} < s) &= \frac{1}{N} \sum_{n=0}^{N-1} \PP(\mathit{\Gamma}_{N, K} < s \mid n_1 = n)\\
&= h_{N-1, K-1}(s),
\end{aligned}
\]
and this completes the proof.
\end{IEEEproof}

\begin{remark}
The deterministic bound given in \eref{bound_Dirac} can be viewed as a simple consequence of our probabilistic analysis. By definition, $h_{N-1, K-1}(s)$ is the $(N-K)$th coefficient of $f_s(x)^K$, a polynomial of degree $(s-1)K$. Now set $s = 2P$. We can see that \eref{bound_Dirac} holds if and only if $N-K$ is larger than the polynomial degree $(2P-1)K$, in which case $h_{N-1, K-1}(2P) = \PP(\mathit{\Gamma}_{N, K} < 2P) = 0$. Consequently, there must exist at least one gap of size greater than or equal to $2P$, and this is all what ProSparse needs to recover $\vx$ successfully.
\end{remark}

\subsection{Success Probabilities of ProSparse}

In what follows, we express the success probabilities of ProSparse in terms of the function $h_{N, K}(s)$ given in \eref{exact_prob}. We consider three different settings as examples.

\begin{example}
$\mPsi$ is an $N \times M$ Vandermonde matrix (for which circularity cannot be exploited) and $\mPhi$ is the identity matrix. Let $\vx = [\vx_1^T, \vx_2^T]^T$, where $\vx_1 \in \C^M$ is $P$-sparse and $\vx_2 \in C^N$ is $K$ sparse. Moreover, the sparsity pattern of $\vx_2$ is drawn uniformly at random using sampling without replacement. In this case, ProSparse can recover $\vx$ if and only if there exists at least one gap of size greater than or equal to $2P$. We thus have the following result:
\[
\PP(\text{ProSparse recovers $\vx$}) = \PP(\mathit{\Delta}_{N, K} \geq 2P) = 1 - h_{N, K}(2P).
\]
\end{example}

\begin{example}
We assume that $\mPsi$ is the same as in the previous example, but $\mPhi$ is a banded matrix with bandwidth $b_{\mPhi}$. Moreover, the diagonal elements of $\mPhi$ are nonzero. We see that $\mathit{\Delta}_{N, K} \ge 2P$ is a necessary condition for ProSparse to recover $\vx$, and $\mathit{\Delta}_{N, K} \ge 2P + 2b_{\mPhi}$ is a sufficient condition. It follows that
\begin{equation}\label{eq:prob_general}
\begin{aligned}
1 - h_{N, K}(2P + 2b_{\mPhi}) \le &\PP(\text{ProSparse recovers $\vx$})\\
&\qquad\qquad \le 1 - h_{N, K}(2P).
\end{aligned}
\end{equation}
\end{example}

\begin{example}
Finally, we consider the case where $\mPsi = \mF_N$ is the $N \times N$ unitary Fourier matrix and $\mPhi = \mI_N$. The problem becomes circulant since the elements in the vector $\vy$ have an underlying periodicity of $N$ samples. Moreover, the problem can also be solved in a dual form
\[
\overline{\mF_N^*\,\vy} = [\mI_N,\mF_N]\,\overline{\vx},
\]
where $\overline{(\cdot)}$ and $(\cdot)^*$ denote the complex conjugate and Hermitian operators, respectively. In the above dual representation, the original spikes become Fourier atoms, and vice versa. Let $\vx = [\vx_1^T, \vx_2^T]^T$ be a $(P, K)$-sparse signal. We assume that the support patterns of $\vx_1$ and $\vx_2$ are drawn independently by sampling without replacement. Exploiting the duality and using the result of Corollary~\ref{corr:exact_proba_circ}, we have
\[
\PP(\text{ProSparse recovers $\vx$}) = 1 - h_{N-1, K-1}(2P)\,h_{N-1, P-1}(2K).
\]
\end{example}

\section{Asymptotic analysis and phase transitions}
\label{sec:phase_trans}

\subsection{Asymptotic Analysis}

\begin{figure}[t]
\centering
\subfigure[$N=128$]{\label{fig:prob_N128}
\includegraphics[width=.42\linewidth]{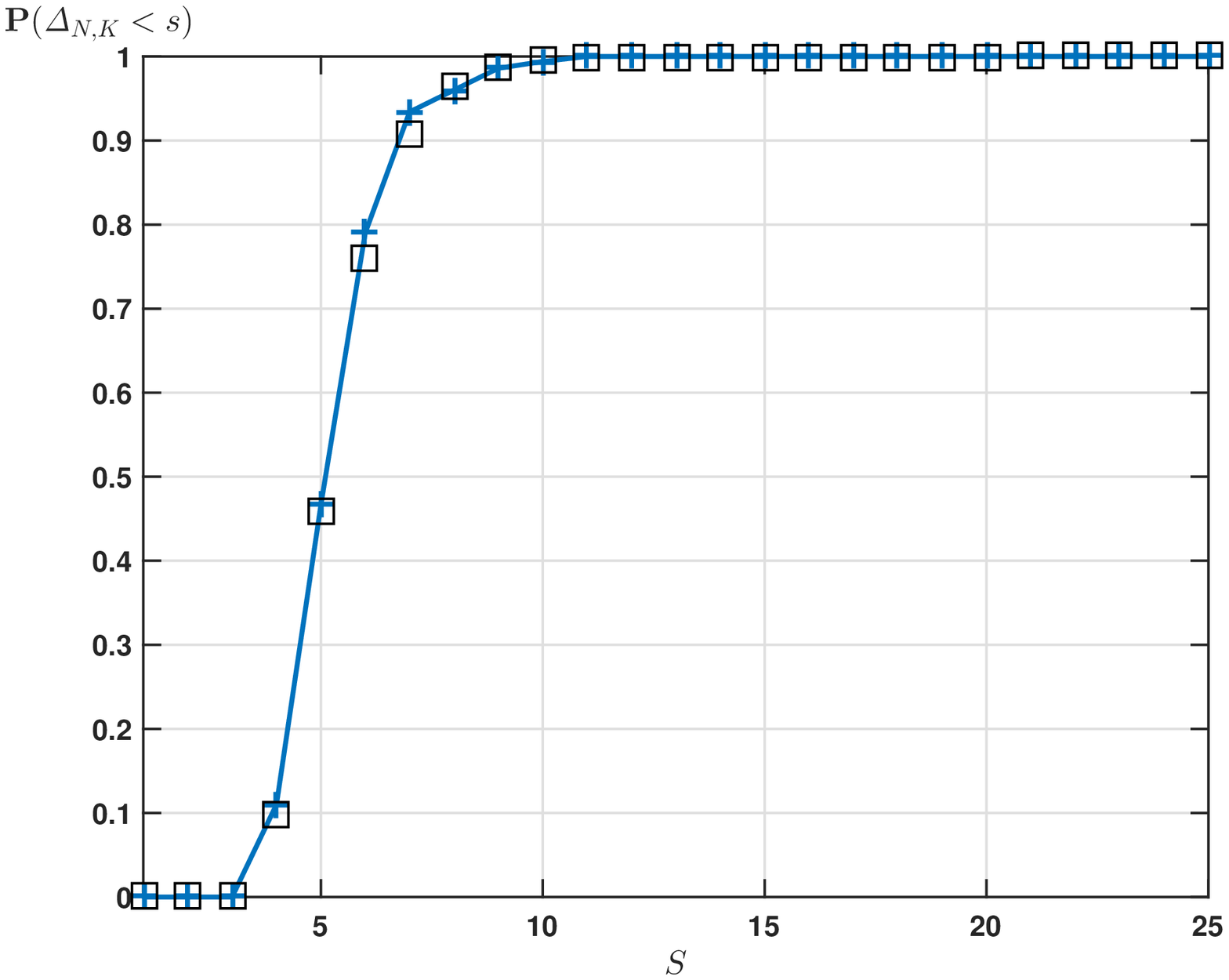}
}
\hspace{2ex}
\subfigure[$N=1024$]{\label{fig:prob_N1024}
\includegraphics[width=.42\linewidth]{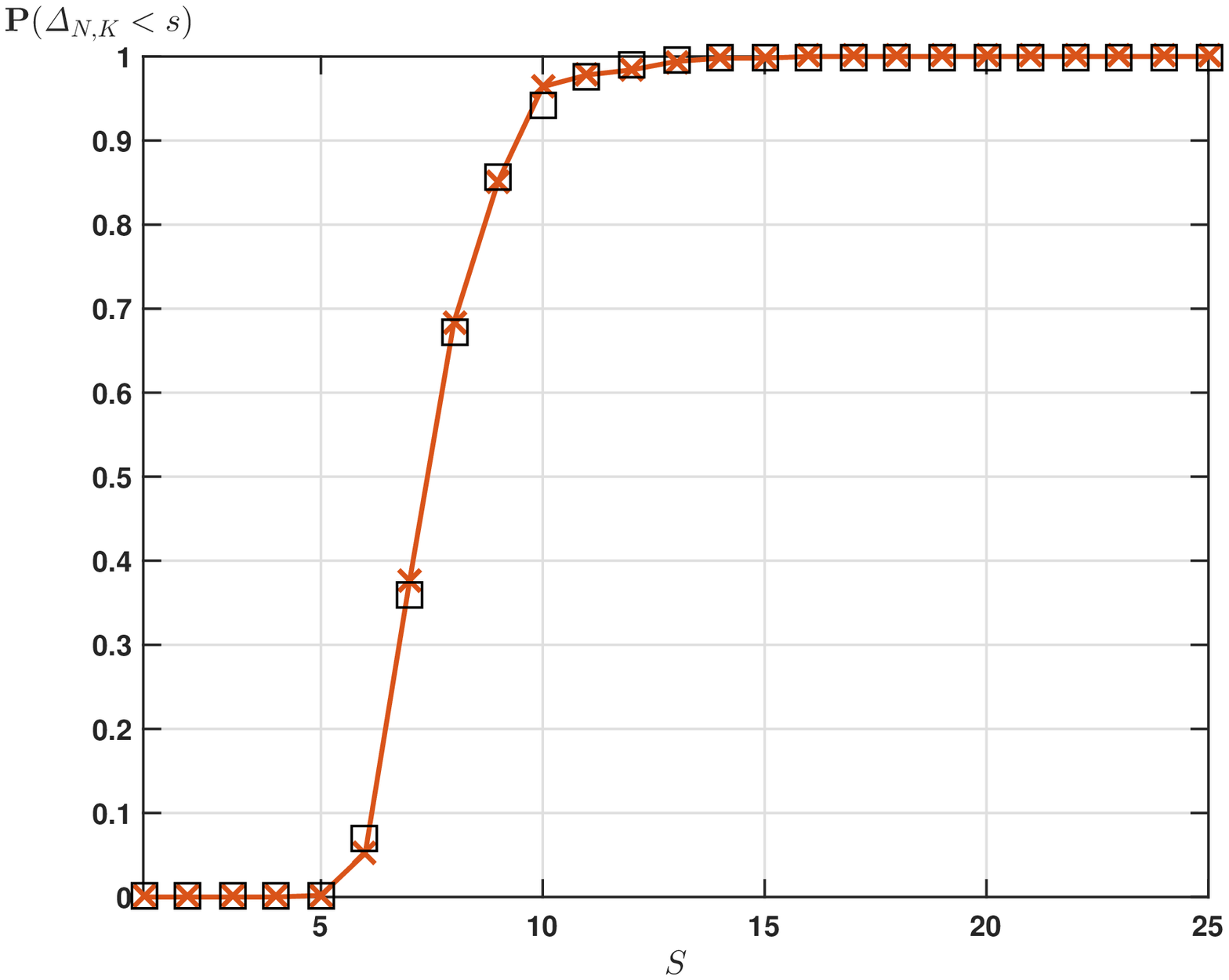}
}
\vspace{1ex}
\subfigure[$N=10^6$]{\label{fig:prob_N1e6}
\includegraphics[width=.42\linewidth]{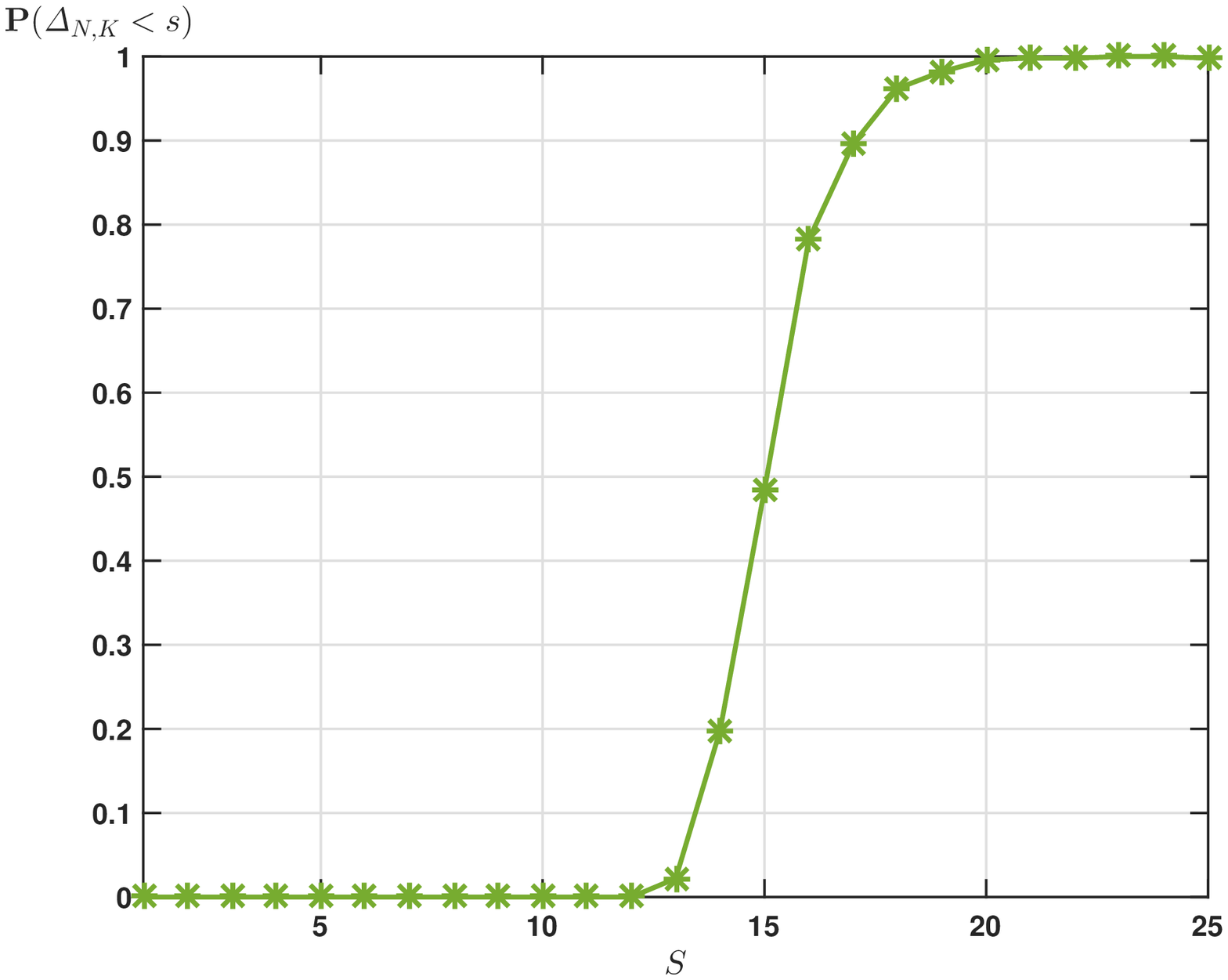}
}
\hspace{2ex}
\subfigure[Rescaled probabilties]{\label{fig:prob_rescaled}
\includegraphics[width=.42\linewidth]{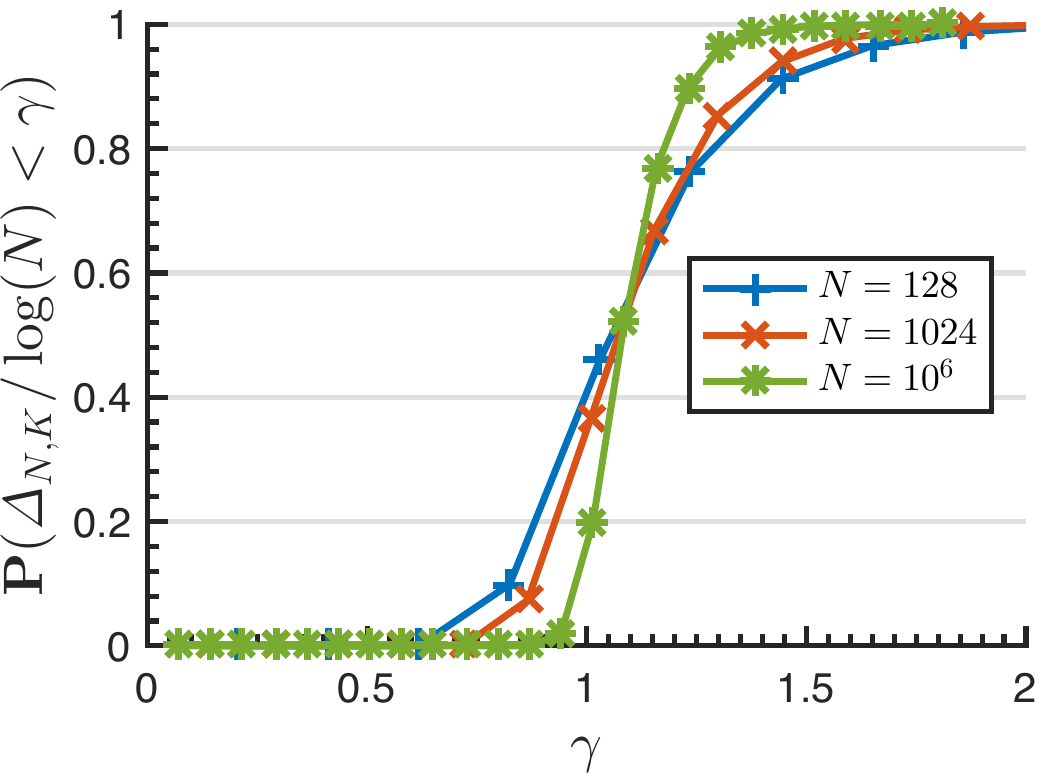}
}
\caption{Illustrations of the probability $\PP(\mathit{\Delta}_{N, K} \le s)$ for different values of $N$, with $K = \lfloor \alpha N \rfloor$ for $\alpha = 0.6$. The exact probabilities shown in (a) and (b) are computed using the expression in \eref{exact_prob}, whereas the estimated probabilities from Monte Carlo simulations are obtained by averaging over $5000$ independent trials.}
\label{fig:exact_vs_monte_carlo}
\end{figure}

Proposition~\ref{prop:exact_proba} provides an analytical formula to compute the success probabilities of the ProSparse algorithm. In Figures~\ref{fig:prob_N128} and \ref{fig:prob_N1024}, we verify the analytical results given in \eref{exact_prob} by comparing them against Monte Carlo simulations, for $N = 128$ and $N = 1024$, respectively. For much larger values of $N$, evaluating the analytical formula in \eref{exact_prob} becomes numerically intractable. Thus, we only present results from Monte Carlo simulations in \fref{prob_N1e6} for $N = 10^6$. In all three experiments, the number of spikes are set to be a constant fraction of $N$, \emph{i.e.}, $K = \lfloor \alpha N \rfloor$ for $\alpha = 0.6$.

Figures~\ref{fig:prob_N128}--\ref{fig:prob_N1e6} show that the probability $\PP(\mathit{\Delta}_{N, K} < s)$ grows rapidly from $0$ to $1$ within a narrow range of $s$, although the point where this transition takes place shifts to the right as $N$ increases. We divide $\mathit{\Delta}_{N, K}$ by $\log(N)$ and overlay the three rescaled probability curves $\PP(\mathit{\Delta}_{N, K}/\log(N) < \gamma)$ as functions of $\gamma$ in \fref{prob_rescaled}. We can see that, after this rescaling, the probability curves are aligned. Moreover, the transitions from regions of low probability to those of high probability become sharper as $N$ increases. This suggests a phase transition phenomenon that takes place in the asymptotic regime where $N \rightarrow \infty$. We confirm and characterize this observation in the following proposition.

\begin{proposition}
\label{prop:main_result}
Let $\vy = [\mPsi, \mPhi] \vx$, where $\mPsi$ is an $N\times M$ Vandermonde matrix with $M \ge N$, $\mPhi$ is an $N \times N$ banded matrix with bandwidth $b_{\mPhi}$, and $\vx = [\vx_1^T, \vx_2^T]^T$ is a $(P, K)$-sparse signal. Furthermore, the support of $\vx_1$ is arbitrary whereas the support of $\vx_2$ is drawn uniformly at random by sampling without replacement. Suppose that the bandwidth $b_{\mPhi}$ is fixed, and let $K = \floor{\alpha N^\delta}$ and $P = \floor{\beta N^{1-\delta} \log N}$ for three constants $0 < \delta \le 1$, $\alpha > 0$ and $\beta > 0$. (When $\delta = 1$, we also require $\alpha < 1$.) It holds that
\begin{equation}
\PP(\text{ProSparse recovers $\vx$}) \xrightarrow[]{N\rightarrow \infty}
\begin{cases}
1, \quad & \text{if } \beta < \beta_c(\alpha, \delta),\\
0, \quad & \text{if } \beta > \beta_c(\alpha, \delta),\\
\end{cases}
\label{eq:phase_transition}
\end{equation}
where the critical threshold is given by
\begin{equation}\label{eq:beta_c}
\beta_c(\alpha, \delta) = \begin{cases}
\frac{\delta}{2 \alpha}, &\text{if } 0 < \delta < 1\\
\frac{-1}{2 \log(1-\alpha)}, &\text{if } \delta = 1.
\end{cases}
\end{equation}
\end{proposition}
\begin{IEEEproof}
We see from \eref{prob_general} that the desired probability is sandwiched between a lower bound and an upper bound. The proof then consists of two parts. The first part deals with the lower bound in \eref{prob_general}. We show that, when $\beta$ is below the phase transition threshold in \eref{phase_transition}, this lower bound converges to $1$ as $N \rightarrow \infty$. The second part of the proof provides an upper bound for $1 - h_{N, K}(2P)$, which is shown to converge to 0 when $\beta$ is above the same threshold. Combined, these two parts verify that a phase transition in the probability of success indeed takes place according to the prediction in \eref{phase_transition}.

\emph{The lower bound:} Let $s$ be a positive integer. We use Proposition~\ref{prop:exact_proba} to write
\begin{equation}\label{eq:prob_lb}
h_{N, K}(s) = \frac{\coef\left\{f_s(x)^{K+1}, x^{N-K}\right\}}{\binom{N}{K}}.
\end{equation}
Since all the coefficients of the polynomial $f_s(x)^{K+1} = (1 + x + \ldots + x^{s-1})^{K+1}$ are positive, it is clear that
\begin{equation}
\coef\left\{f_s(x)^{K+1}, x^{N-K}\right\} \le \frac{f_s(x)^{K+1}}{x^{N-K}}
\end{equation}
for all $x > 0$. Replacing $f_s(x)$ by its equivalent expression $(1-x^s)/(1-x)$ and setting $x = 1 - K/N$, we then get
\begin{equation}
\coef\left\{f_s(x)^{K+1}, x^{N-K}\right\} \le \frac{\big(1-(1-K/N)^s\big)^{K+1}}{(1-K/N)^{N-K}\,(K/N)^{K+1}}.\label{eq:num_bnd}
\end{equation}

Using the following standard estimate on factorials
\[
\sqrt{2\pi} \, n^{n+1/2} e^{-n} \le n! \le n^{n+1/2} e^{1-n},
\]
which holds for all positive integers $n$, we can verify that
\begin{equation}
\binom{N}{K} \ge C K^{-1/2} \frac{N^N}{K^K(N-K)^{N-K}},
\label{eq:binom_low}
\end{equation}
where $C$ is a constant that does not depend on $N$ or $K$. Substituting \eref{num_bnd} and \eref{binom_low} into \eref{prob_lb} leads to
\begin{equation}\label{eq:h_bnd}
h_{N, K}(s) \le C N \big(1 - (1-K/N)^s\big)^{K+1}.
\end{equation}

Let $\beta_1, \beta_2$ be two arbitrary constants such that
\begin{equation}\label{eq:threshold}
0 < \beta_1 < \beta_2 < \beta_c(\alpha, \delta),
\end{equation}
where $\beta_c(\alpha, \delta)$ is the critical threshold given in \eref{beta_c}. We show next that, for $P = \floor{\beta_1 N^{1-\delta} \log N}$ and $s = 2 P + 2 b_{\mPhi}$, the right-hand side of \eref{h_bnd} goes to $0$ as $N \rightarrow \infty$. To that end, we first note that $s < 2 \beta_2 N^{1-\delta} \log N$ for all sufficiently large $N$. Substituting this bound into the right-hand side of \eref{h_bnd} gives us
\begin{align}
&h_{N, K}(s) \le C N (1 - (1 - K/N)^{2 \beta_2 N^{1-\delta} \log N})^{K+1}\nonumber\\
		&\qquad= CN \exp\big\{(K+1)\log\big(1-N^{2\beta_2 N^{1-\delta} \log(1-K/N)}\big)\big\}\nonumber\\
&\qquad\le CN \exp\big\{-\alpha N^{\delta + 2\beta_2 N^{1-\delta} \log(1-K/N)}\big\} \label{eq:h_bnd1},
\end{align}
where \eref{h_bnd1} follows from the inequality $\log(1+x) \le x$ which holds for all $x > -1$ and from the fact that $K+1 > \alpha N^\delta$. For any $0 < \delta \le 1$, it is easy to verify that
\[
\begin{aligned}
\lim_{N \rightarrow \infty} (\delta + 2 \beta_2 N^{1-\delta} \log(1-K/N)) &= \delta\left(1 - \beta_2/ \beta_c(\alpha, \delta)\right)\\
&> 0,
\end{aligned}
\]
where the inequality is due to \eref{threshold}. It then follows from \eref{h_bnd1} that $h_{K, N}(2P + 2 b_{\mPhi}) \xrightarrow[]{N\rightarrow \infty} 0$. This establishes the first part of \eref{phase_transition}.

\emph{Upper bound:} It remains to show the second part of \eref{phase_transition}. To do so, we consider the upper bound in \eref{prob_general}. Let $s = 2 P$, where $P = \floor{\beta_3 N^{1-\delta}\log N}$ for some fixed constant
\begin{equation}\label{eq:threshold2}
\beta_3 > \beta_c(\alpha, \delta).
\end{equation}
A simple union bound yields
\begin{align}
1 - h_{N, K}(s) &= \PP(\mathit{\Delta}_{N, K} \geq s)\\
& = \PP\big( \cup_{k=0}^{K} \{d_k \geq s\}\big) \label{eq:prob_set_union}\\
& \leq \sum_{k=0}^{K} \PP(d_k \geq s),   \label{eq:union_bound}
\end{align}
where $\set{d_k}$ are the gaps defined in \eref{gaps}. Due to symmetry, the random variables $\set{d_k}$ are exchangeable, and thus they have the same marginal distributions. It follows that 
\begin{equation}\label{eq:sum_prob}
\sum_{k=0}^{K} \PP(d_k \ge s) = (K+1) \, \PP(d_0 \geq s).
\end{equation}
For any integer $s \in [0, N-K]$, the probability $\PP(d_0 \geq s)$ is given by $\binom{N-s}{K} \big/ \binom{N}{K}$, since the event of having the first gap of size at least $s$ is equivalent to having the 
$K$ breaking points all appearing in the last $N-s$ locations. Substituting this expression into \eref{sum_prob}, we get
\begin{align}
&\sum_{k=0}^{K} \PP(d_k \geq s) =  (K+1) \, \frac{(N-s)!\,(N-K)!}{N!\,(N-K-s)!}\\
&\quad\qquad= (K+1) \, \frac{(N-K)\ldots(N-K-s+1)}{N(N-1)\ldots(N-s+1)},\label{eq:marginal}
\end{align}
for $s \le N - K$. Next, we consider the two cases $0 < \delta < 1$ and $\delta = 1$ separately.


Case 1: $0 < \delta < 1$. From \eref{marginal},
\begin{align}
\sum_{k=0}^{K} \PP(d_k \geq s) &= (K+1) \frac{(N-K)^s}{N^s} \prod_{i = 0}^{s-1} \frac{1 - i/(N-K)}{1 - i/N}\\
&\le (K+1) (1 - K/N)^s \label{eq:prod_ineq}\\
&\le \frac{K+1}{N^\delta} N^\delta \exp\{-s K / N\}, \label{eq:c1}
\end{align}
where \eref{prod_ineq} and \eref{c1} follow from the inequalities $i/(N-K) \ge i/N$ and $\log(1+x) \le x$ for all $x > 1$, respectively. Substituting $s = 2 \floor{\beta_3 N^{1-\delta}\log N}$ and $K = \floor{\alpha N^\delta}$ into \eref{c1} gives us
\begin{equation}\label{eq:bound_c1}
\sum_{k=0}^{K} \PP(d_k \geq s) \le (\alpha + 1) N^{\delta - 2 \alpha \beta_3},
\end{equation}
for all sufficiently large $N$. Since $\beta_c(\alpha, \delta) = \delta/(2\alpha)$ when $\delta < 1$, the condition \eref{threshold2} implies that the right-hand side of \eref{bound_c1} tends to zero as $N \rightarrow \infty$. The second part of \eref{phase_transition} for the case of $0 < \delta < 1$ then follow from \eref{union_bound} and \eref{prob_general} immediately.

Case 2: $\delta = 1$. We first rewrite \eref{marginal} as
\begin{equation}
\sum_{k=0}^{K} \PP(d_k \geq s) =  (K+1) \prod_{i=0}^{s-1} \frac{1-\frac{K}{N}- \frac{i}{N}}{1 - \frac{i}{N}}.
\end{equation}
To bound the above expression, we introduce a function $r(x) \bydef \log \frac{\eta - x}{1-x}$, where $\eta$ is a constant satisfying
\begin{equation}\label{eq:eta}
 1 - \alpha < \eta < e^{-1/(2\beta_3)}.
\end{equation}
Note that, due to \eref{threshold2}, the above interval is nonempty and we can always find a suitable $\eta$. Since $K/ N \rightarrow \alpha$, it holds for all sufficiently large values of $N$ that
\begin{align}
\sum_{k=0}^{K} \PP(d_k \geq s) &\le (K+1) \exp\Big\{\sum_{i=0}^{s-1} r({i}/{N})\Big\}\\
&\le (K+1) \exp\Big\{ N \int_0^{\frac{s-1}{N}} r(x) dx\Big\}, \label{eq:upper_exps_int}
\end{align}
where the second inequality is due to the fact that the function $r(x)$ is negative and monotonically decreasing on the interval $x \in [0, \eta)$.

Let $g(x) \bydef \int_{0}^{x} r(y)\,dy$. It is easy to verify that $g(x)$ is a smooth function in a neighborhood around $x = 0$. Moreover, $g'(0)=\log \eta$ and 
$g''(0)=1-\frac{1}{\eta}$. Recall that $s = 2 P = 2 \floor{\beta_3 \log N}$ and thus $(s-1)/N = o(1)$. A Taylor expansion at $x=0$ leads to
\begin{equation}
g\Big(\frac{s-1}{N}\Big) = (\log \eta) \, \frac{s-1}{N} + \frac{\eta-1}{2\eta}\,\frac{(s-1)^2}{N^2}\,(1+o(1)).
\label{eq:g_taylor}
\end{equation}
Inserting \eqref{eq:g_taylor} into \eqref{eq:upper_exps_int} and setting $s = 2 \floor{\beta_3 \log N}$, we get
\begin{align}
\sum_{k=0}^{K} \PP(d_k \geq s) &\leq \frac{K+1}{\eta} \exp\{2\beta_3 \log \eta \log N\} (1 + o(1))\\
&= \frac{\alpha}{\eta} N^{1 + 2\beta_3 \log \eta} (1+o(1)).
\end{align}
By the second inequality in \eref{eta}, the exponent $1 + 2\beta_3 \log \eta < 0$. Thus, the above upper bound tends to $0$ as $N \rightarrow \infty$. Now applying \eref{union_bound} and \eref{prob_general}, we have shown the second part of \eref{phase_transition} for the case of $\delta = 1$.
\end{IEEEproof}

%

\subsection{Comparison with Basis Pursuit}
\label{sec:comparison}

A popular way to solve the sparse representation problem discussed in this paper is to use basis pursuit (BP),
namely, by solving:
\begin{equation}
\text{arg}\,\min_{\widetilde{\vx}} \| \widetilde{\vx} \|_1 \hspace{5mm}
\mbox{s.t.} \hspace{5mm} \vy=\mD \widetilde{\vx}.
\label{eq:BP_problem}
\end{equation}
The performance of BP is well understood and, in particular, probabilistic recovery guarantees for BP have been derived in \cite{CandesR:2005,Tropp:2008,Kuppinger:2012}.

Let $\mD$ be a dictionary with column vectors $\set{\vd_i}$. Its mutual coherence is defined as
\[
\mu(\mD) \stackrel{\mbox{def}}{=}  \max_{k \neq \ell} \frac{|\vd_k^H \vd_\ell |}{\|\vd_k \| \| \vd_\ell \|}.
\]
The case when $\mD =  [\mPsi, \mPhi]$ is the union of two orthonormal bases in $\C^N$ was considered in \cite{CandesR:2005,Tropp:2008}. Let $\vx \in \C^{2N}$ be a $(P, K)$-sparse signal, with an arbitrarily-chosen sparsity pattern of $P$ nonzero coefficients corresponding to $\mPsi$ and a randomly-chosen sparsity pattern of $K$ nonzero coefficients corresponding to $\mPhi$. It has been shown that BP can recover $\vx$ from $\vy = \mD \vx$ with high probability if 
\begin{equation}
P + K < C \frac{\mu(\mD)^{-2}}{\log(2N)}
\label{eq:BP_AverageBound}
\end{equation}
for some constant $C$. Later, this result was extended in \cite{Kuppinger:2012, Pope:2013} to more general cases where $\mD$ is the union of two arbitrary sub-dictionaries. It was shown in \cite[Theorem 6]{Kuppinger:2012} that BP can recover $\vx$ with high probability if, in addition to \eref{BP_AverageBound}, the sparsity levels $P$ and $K$ satisfy some extra constraints given by the mutual coherences computed on the sub-dictionaries $\mPsi$ and $\mPhi$. This recovery guarantee is further improved in \cite{Pope:2013}, where the authors exploit the individual coherence parameters of the sub-dictionaries as well as the mutual coherence between $\mPsi$ and $\mPhi$. (See \cite[Theorem 5]{Pope:2013} for details.)

To compare the probabilistic ProSparse bound given in Proposition~\ref{prop:main_result} with the above BP bounds, we consider the special case when the sub-dictionary $\mPhi \in \R^{N \times N}$ is the identity matrix and the sub-dictionary $\mPsi \in \C^{N \times M}$ (with $M \ge N$) is a Fourier frame with entries
\[
\Psi_{n, m} = \frac{1}{\sqrt{N}} e^{-j 2\pi mn/M}.
\]
When $M = N$, both sub-dictionaries are orthonormal bases, and the mutual coherence of the entire dictionary becomes $\mu(D) = 1/\sqrt{N}$. It follows that the BP bounds in \eref{BP_AverageBound} becomes
\begin{equation}\label{eq:BP_average}
P + K < C \frac{N}{\log N}.
\end{equation}
In comparison, the result of Proposition~\ref{prop:main_result} shows that ProSparse can recover $\vx$ with high probability if
\begin{equation}\label{eq:ProSparse_average}
K \le \lfloor \alpha N^\delta \rfloor \quad \text{and}\quad P < \beta_c(\alpha, \delta) N^{1-\delta} \log N
\end{equation}
for $\delta \in (0, 1]$ and $\alpha \in (0, 1)$. ProSparse requires an unbalanced distribution of atoms between the local basis $\mPhi$ and 
the Fourier basis $\mPsi$. Moreover, the requirement that $P$ is of order $N^{1-\delta} \log N$ is much worse than what the BP bounds in \eref{BP_average} allows.

The situation becomes different when $\mPsi$ is a \emph{redundant} Fourier frame, \emph{i.e.}, when $M > N$. It is easy to show that the mutual coherence parameters in this case satisfy
\begin{equation}\label{eq:mutual}
\mu(\mD) \ge \mu(\mPsi) \ge \frac{1}{N} \left\vert\sum_{n=0}^{N-1} e^{j2\pi n/M}\right\vert = \frac{1}{N}\left\vert\frac{1-e^{j2\pi N/M}}{1-e^{j2\pi/M}}\right\vert,
\end{equation}
where $\mu(\mPsi)$ is the mutual coherence parameter of the sub-dictionary $\mPsi$. Let $d = M/N$ be the redundancy factor. Fixing $d > 1$ and letting $N \rightarrow \infty$, the right-hand side of \eref{mutual} converges to a constant
\[
\frac{d \sqrt{1-\cos(2\pi/d)}}{\sqrt{2}\pi}
\]
that does not depend on $N$. This implies that $\mu(\mPsi) = \mathcal{O}(1)$. The BP recovery guarantee given in \cite[Theorem 5]{Pope:2013} requires, among other conditions, that $P \mu^2(\mPsi) < C$ for some constant $C$, and thus, $P$ needs to be of $\mathcal{O}(1)$. In contrast, since the ProSparse bound does not depend on the mutual coherence, the same bound in \eref{ProSparse_average} still holds even when $\mPsi$ is very redundant. This suggests that ProSparse might have an advantage over BP when dealing with redundant sub-dictionaries.

We need to highlight at this point that, although the ProSparse bound is asymptotically tight, the same cannot be claimed for the existing BP bounds. They are merely sufficient conditions for BP to succeed. The actual performance of BP can be better than what the bounds predict. Nevertheless, we observe from numerical simulations that ProSparse can indeed outperform BP when $\mPsi$ is redundant. To illustrate this point, we show in \fref{ProSparse_BP} the comparison of ProSparse and BP for sparse recovery from a dictionary consisting of a Fourier frame and a canonical basis. The Fourier frame has size $N\times M$ with $N = 128$ and $M = d N$, where $d$ is the redundancy factor. We consider three settings: $d = 4, 8$ and $16$, corresponding to increasingly redundant frames. Empirical success probabilities of both algorithms are computed by generating 50 different realizations of the sparse vector $\vx$ for each sparsity combination $(P, K)$. The locations of the nonzero entries of $\vx$ are drawn uniformly at random from the sets $\{ 1,2,...,M\} $ and $\{1,2,...,N \}$. The amplitudes of these elements are complex, with the real and imaginary parts drawn from $\mathcal{N}(0,1)$.

The BP results in \fref{ProSparse_BP:1}--\fref{ProSparse_BP:3} are obtained by solving the $\ell_1$-minimization problem in \eref{BP_problem} with CVX, a package for specifying and solving convex programs \cite{cvx, gb:08}. The success of the algorithm is measured by computing $\| \tilde{\vx}-\vx \|^2/ \| \vx \|^2$ and by checking that it is below  $10^{-3}$. Here $\tilde{\vx}$ is the reconstructed sparse vector. The ProSparse results in \fref{ProSparse_BP:4} are obtained by checking that the maximum gap between consecutive spikes is at least equal to $2P$. When this is the case, the Fourier atoms can be recovered and therefore perfect reconstruction of the vector $\vx$ is achieved. Since the performance of ProSparse does not change with the redundancy factor $d$ of the Fourier frame, only one figure is shown for ProSparse.

\begin{figure}[t]
\centering
\subfigure[Basis pursuit ($d=4$)]{\label{fig:ProSparse_BP:1}
\includegraphics[width=.42\linewidth]{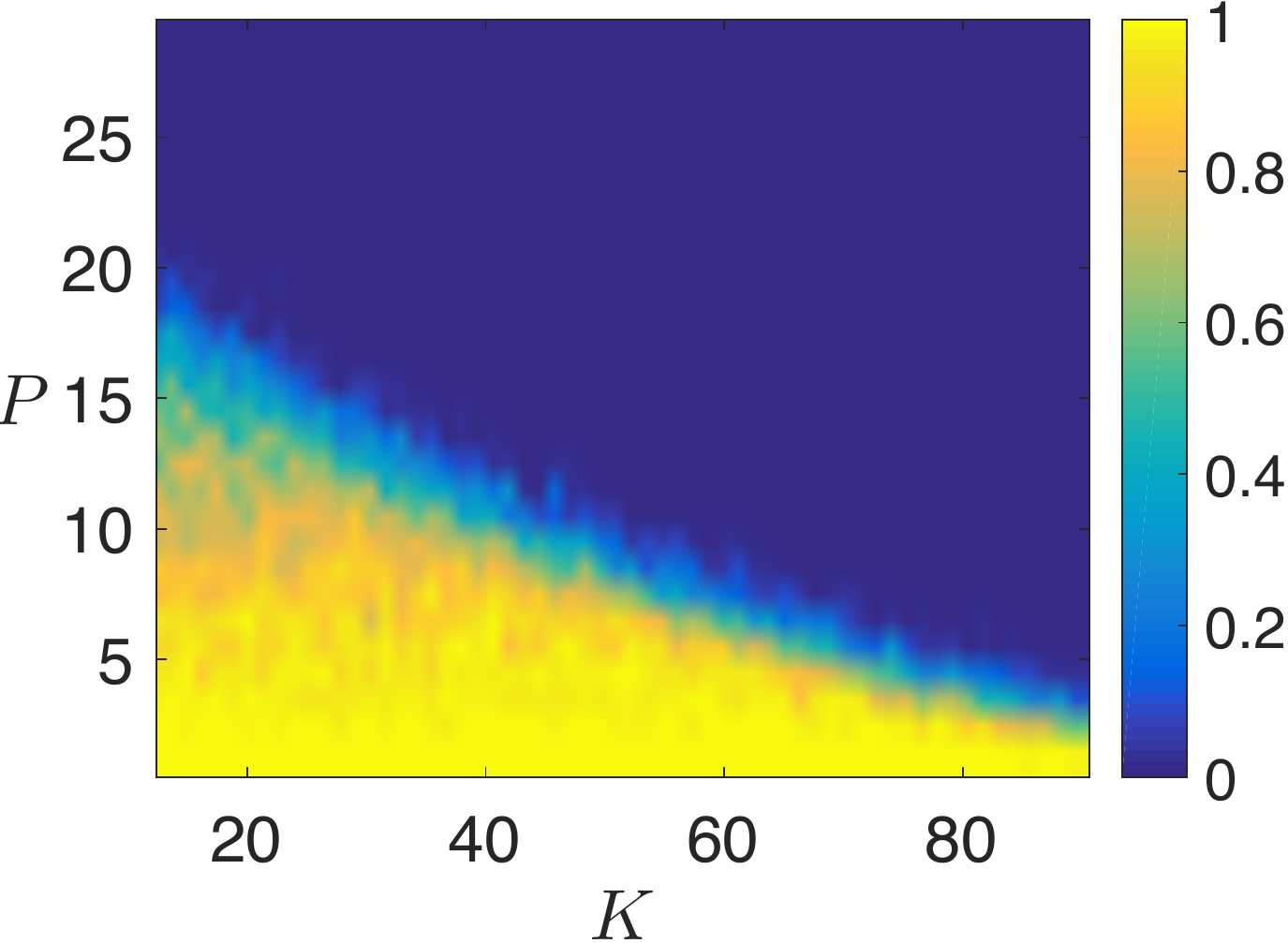}
}
\subfigure[Basis pursuit ($d=8$)]{\label{fig:ProSparse_BP:2}
\includegraphics[width=.42\linewidth]{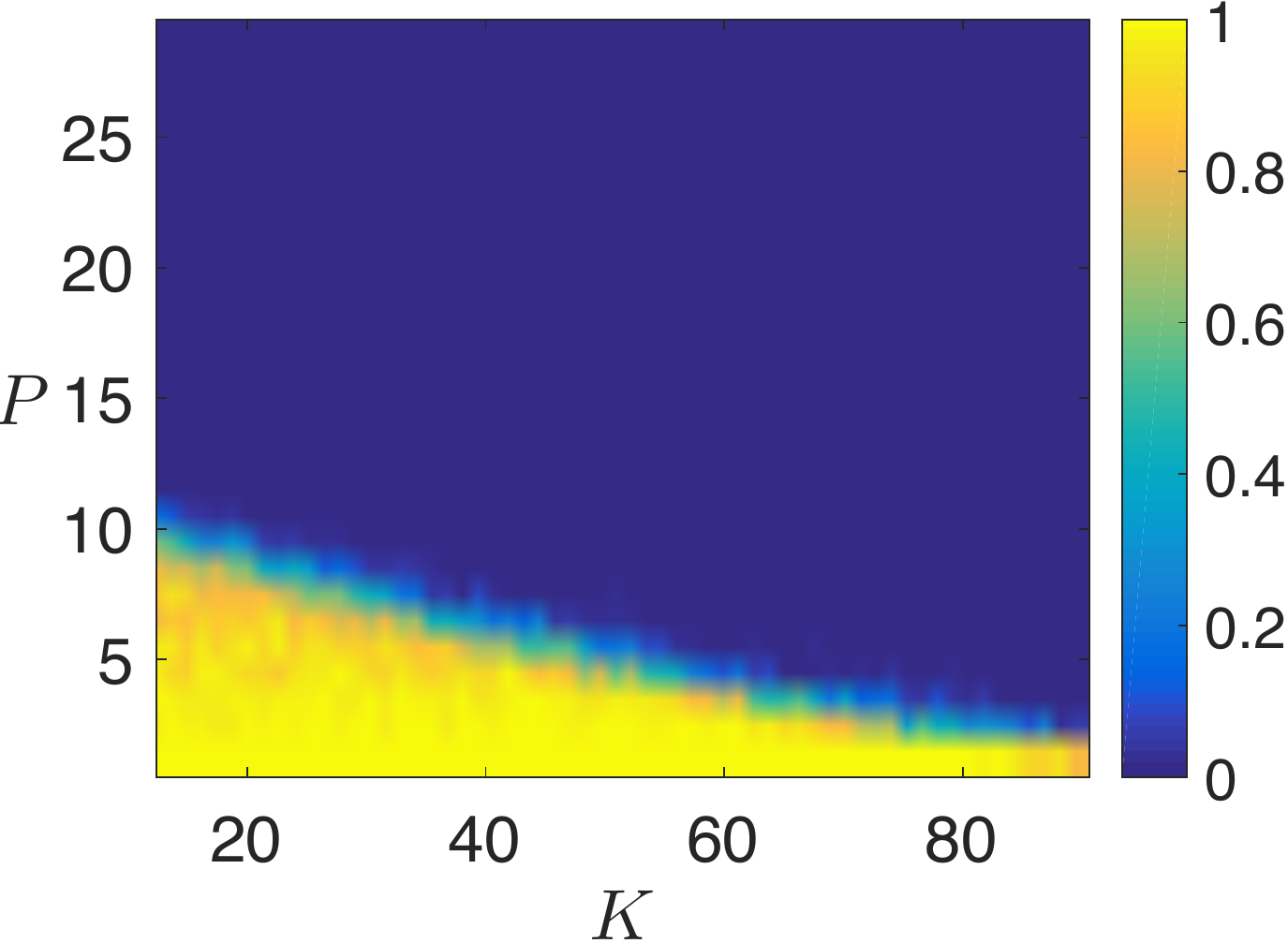}
}
\vspace{1ex}
\subfigure[Basis pursuit ($d=16$)]{\label{fig:ProSparse_BP:3}
\includegraphics[width=.42\linewidth]{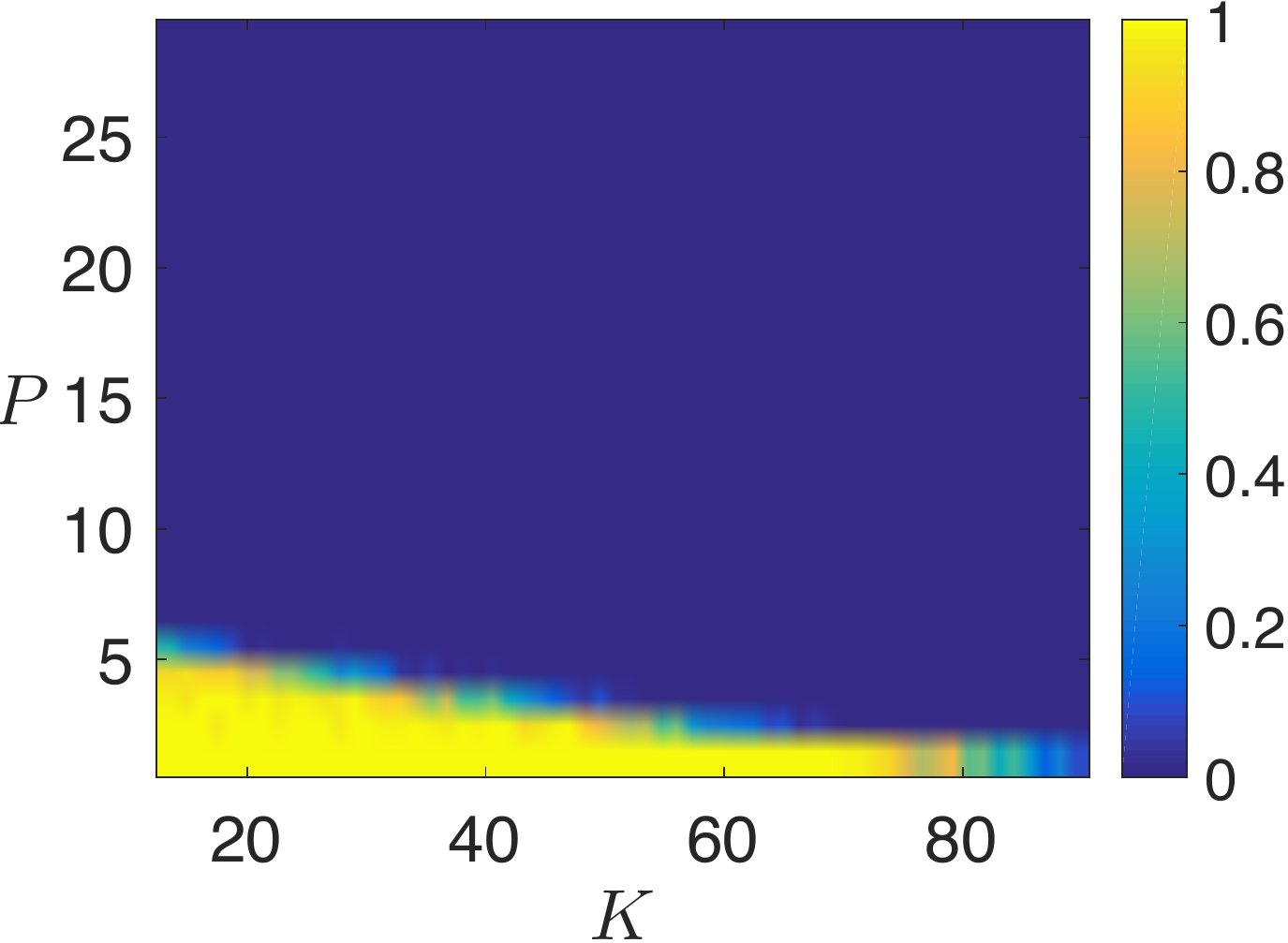}
}
\subfigure[ProSparse]{\label{fig:ProSparse_BP:4}
\includegraphics[width=.42\linewidth]{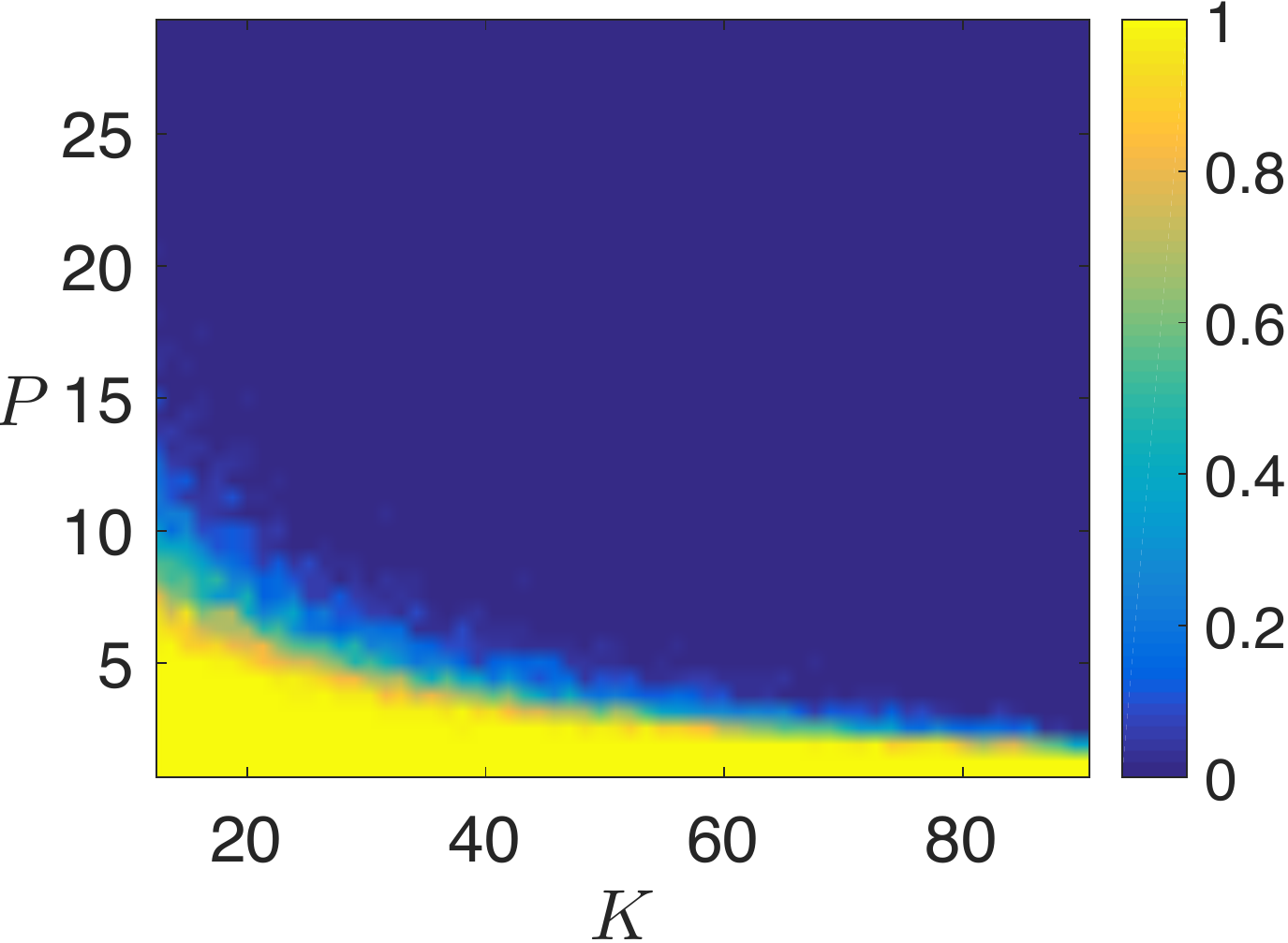}
}
\caption{Comparison between ProSparse and basis pursuit for sparse recovery from a dictionary consisting of a Fourier frame and a canonical basis. The Fourier frame is of size $N \times M$, with $N = 128$. We test three redundancy factors: $d = M/N$ is set to $4, 8$ and $16$, respectively. Shown in the figures are the empirical success probabilities of the two algorithms. Since the performance of ProSparse does not change with the redundancy factor $d$, only one figure is shown for ProSparse.
}
\label{fig:ProSparse_BP}
\end{figure}

As expected, both algorithms show a phase transition behavior since the $(K, P)$-plane is clearly split in two regions, one where the algorithms achieve perfect reconstruction with very high probability, and the other where the algorithms fail in most of the cases. BP outperforms ProSparse when $d = 4$. However, as we increase the redundancy ratio, which leads to increased mutual coherence between the dictionary elements, the performance of BP degrades. In contrast, the performance of ProSparse is not affected by the increasing redundancy. At $d = 16$, ProSparse outperforms BP by recovering sparse vectors in a larger region of the $(K,P)$-plane. This suggests that, in the case where $\mPhi$ is a highly redundant frame, ProSparse can be the method of choice.

\section{Conclusions}
\label{sec:conclusion}

We presented an average-case performance analysis of ProSparse to recover a sparse signal from the union of a Vandermonde matrix and a banded matrix. The underlying probability model turns out to be equivalent to the problem of discrete circle covering. Based on a simple generating-function approach, we presented a new analytical expression for the exact success probabilities of ProSparse. We also analyzed the above-mentioned success probabilities in the high-dimensional regime. Our asymptotic analysis reveals a phase transition phenomenon regarding the performance of ProSparse. Unlike BP, the average-case performance of ProSparse does not depend on the mutual coherence of the underlying dictionary. This unique property makes ProSparse a potentially more attractive choice than BP when the sub-dictionary corresponding to the Vandermonde matrix is highly redundant.

\bibliographystyle{IEEEtran}
\bibliography{refs}

%
%

\end{document}

%% file: commands.tex
\usepackage{amsmath}
\usepackage{amssymb}
\usepackage{latexsym}
\usepackage{verbatim}
\usepackage{subfigure}
\usepackage[final]{graphicx}
\usepackage{psfrag}

\newtheorem{proposition}{Proposition}

\newtheorem{corollary}{Corollary}
\newtheorem{remark}{Remark}

\newtheorem{example}{Example}

{\begin{list}               
    {$\bullet$ \hfill}{
        \setlength{\leftmargin}{\parindent}
        \setlength{\parsep}{0.04\baselineskip}
        \setlength{\itemsep}{0.5\parsep}
        \setlength{\labelwidth}{\leftmargin}
        \setlength{\labelsep}{0em}}
    }
{\end{list}}

\providecommand{\eref}[1]{\eqref{eq:#1}}  
\providecommand{\cref}[1]{Chapter~\ref{chap:#1}}
\providecommand{\sref}[1]{Section~\ref{sec:#1}}
\providecommand{\fref}[1]{Figure~\ref{fig:#1}}

\providecommand{\R}{\ensuremath{\mathbb{R}}}
\providecommand{\C}{\ensuremath{\mathbb{C}}}

\providecommand{\set}[1]{\left\{#1\right\}}

\providecommand{\bydef}{\overset{\text{def}}{=}}

\renewcommand{\vec}[1]{\ensuremath{\boldsymbol{#1}}}
\providecommand{\mat}[1]{\ensuremath{\boldsymbol{#1}}}

 \providecommand{\mD}{\mat{D}}
\providecommand{\mF}{\mat{F}}
\providecommand{\mI}{\mat{I}}

\providecommand{\mPhi}{\mat{\Phi}}
\providecommand{\mPsi}{\mat{\Psi}}

 \providecommand{\vd}{\vec{d}}

\providecommand{\vr}{\vec{r}}

\providecommand{\vx}{\vec{x}} \providecommand{\vy}{\vec{y}}